\documentclass[10pt,conference]{IEEEtran}

\IEEEoverridecommandlockouts

\usepackage{multirow}
\usepackage{subcaption}
\usepackage{enumitem}
\usepackage{algorithm}
\usepackage[ruled,linesnumbered,algo2e]{algorithm2e}
\usepackage{tcolorbox}
\usepackage{listings}
\usepackage{booktabs}
\usepackage{graphicx} 
\usepackage{bm}
\usepackage{xspace}
\usepackage{caption}
\usepackage{url}
\usepackage{amsmath,amssymb,amsfonts}
\usepackage{textcomp}
\usepackage{xcolor}
\usepackage{cite}
\usepackage{tcolorbox}
\usepackage{color,xcolor,colortbl}
\def\BibTeX{{\rm B\kern-.05em{\sc i\kern-.025em b}\kern-.08em
    T\kern-.1667em\lower.7ex\hbox{E}\kern-.125emX}}

\DontPrintSemicolon
\SetAlgoNlRelativeSize{0}
\SetAlgoNlRelativeSize{-1}

\definecolor{mygray}{gray}{.9}

\newcommand{\tool}{RepoSPD\xspace}
\newcommand{\graph}{RepoCPG\xspace}
\newcommand{\FuseCPG}{MergeCPG\xspace}

\newcommand{\http}{\url{https://github.com/Xin-Cheng-Wen/RepoSPD}}

\usepackage{tikz}
\newcommand*{\circled}[1]{\lower.7ex\hbox{\tikz\draw (0pt, 0pt)%
    circle (.5em) node {\makebox[1em][c]{\small #1}};}}

\begin{document}

\title{Repository-Level Graph Representation Learning for Enhanced Security Patch Detection}

\author{\IEEEauthorblockN{Xin-Cheng Wen$^{1}$, Zirui Lin$^{1}$, Cuiyun Gao$^{1,2\ast}$, Hongyu Zhang$^{3}$, Yong Wang$^{4}$, Qing Liao$^{1,2}$}

\IEEEauthorblockA{$^1$ Harbin Institute of Technology, Shenzhen, China}
\IEEEauthorblockA{$^2$ Peng Cheng Laboratory, Shenzhen, China}

\IEEEauthorblockA{$^3$ Chongqing University, Chongqing, China}

\IEEEauthorblockA{$^4$ Anhui Polytechnic University, Anhui, China}
\IEEEauthorblockA{xiamenwxc@foxmail.com, 210110128@stu.hit.edu.cn, gaocuiyun@hit.edu.cn, \\ hyzhang@cqu.edu.cn, yongwang@ahpu.edu.cn, liaoqing@hit.edu.cn}

\thanks{$^{\ast}$ Corresponding author.}}

\maketitle

\begin{abstract}
Software vendors often silently release security patches without providing sufficient advisories (e.g., Common Vulnerabilities and Exposures) or delayed updates via resources (e.g., National Vulnerability Database). 
Therefore, it has become crucial to detect these security patches to ensure secure software maintenance. However, existing methods face the following challenges: (1) They primarily focus on the information within the patches themselves, overlooking the complex dependencies in the repository. (2) Security patches typically involve multiple functions and files, increasing the difficulty in well learning the representations.
To alleviate the above challenges, this paper proposes a \textit{Repo}sitory-level {S}ecurity {P}atch {D}etection framework named \textit{\tool}, which comprises three key components: 
1) a repository-level graph construction, \graph, which represents software patches by merging pre-patch and post-patch source code at the repository level; 
2) a structure-aware patch representation, which fuses the graph and sequence branch and aims at comprehending the relationship among multiple code changes; 
3) progressive learning, which facilitates the model in balancing semantic and structural information.
To evaluate \tool, we employ two widely-used datasets in security patch detection: SPI-DB and PatchDB. We further extend these datasets to the repository level, incorporating a total of 20,238 and 28,781 versions of repository in C/C++ programming languages, respectively, denoted as SPI-DB* and PatchDB*.
We compare \tool with six existing security patch detection methods and five static tools. Our experimental results demonstrate that \tool outperforms the state-of-the-art baseline, with improvements of 11.90\%, and 3.10\% in terms of accuracy on the two datasets, respectively. These results underscore the effectiveness of \tool in detecting security patches.
Furthermore, \tool can detect 151 security patches, which outperforms the best-performing baseline by 21.36\% with respect to accuracy.


\end{abstract}

\pagestyle{plain}

\maketitle
\section{Introduction}\label{sec:intro} 
In recent years, the increasing number and diversity of vulnerabilities~\cite{Statista1, DBLP:journals/tse/WenGLWLL24} in Open-Source Software (OSS) have presented significant challenges to software security, posing substantial risks to society~\cite{loss,DBLP:conf/kbse/WenWGWLG23}. According to the Synopsys~\cite{Synopsys} report in 2024, 84\% of codebases contain at least one open-source vulnerability, and 91\% of these codebases include components that are outdated by ten or more versions~\cite{SynopsysReport}.
There is a critical need for the timely detection of software security patches to mitigate 
attacks~\cite{DBLP:conf/dsn/Wang0BJ19}. However, the management of security patches is often subjective by managers~\cite{tamjidyamcholo2022subjectivity, DBLP:conf/dsn/Wang0BJ19}, leading software vendors to release security updates without sufficient
publicity~\cite{DBLP:conf/ccs/LiP17}. This practice of silently releasing patches complicates the identification and remediation processes, as users or administrators are frequently overwhelmed by the growing number of patches~\cite{DBLP:conf/kbse/MirhosseiniP17}, which often results in delayed software updates and vulnerability reports. 
The existing study has revealed that over 82\% of user-submitted software vulnerability reports are filed more than 30 days after the initial detection~\cite{DBLP:conf/icsm/ThungLJLRD12}.
For example, CVE-2024-24919~\cite{CVE-2024-24919}, an information disclosure vulnerability, was first disclosed on May 24th, 2024. However, threat actors had begun exploiting this vulnerability as early as April 30th, targeting over 51 IP addresses~\cite{51ip}. Consequently, more than a hundred thousand users, including those in banks, federal agencies, and large enterprises, faced significant exposure risks.
Therefore, it is imperative for both users and developers to automatically distinguish security patches from other updates and prioritize those that directly address security vulnerabilities. 

Deep Learning (DL)-based methods have achieved great success in identifying security patches as they can reduce the dependence on the quality of commit messages in patches and offer a broader spectrum of capabilities in detecting various security patches~\cite{cheng2022bug, DBLP:journals/caaitrit/ZhangLZJZ23,DBLP:journals/fcsc/LiuZSTZ24}. Current methods can be categorized into sequence-based and graph-based approaches.
Sequence-based methods process the sequential inputs of all code changes in a patch and then utilize DL models to determine whether a code commit fixes a vulnerability. For example, PatchRNN~\cite{DBLP:conf/milcom/WangWFSJBG21/patchrnn} uses both commit messages and code changes as input and then employs the
Recurrent Neural Network (RNN)~\cite{rnn} to deal with the input sequentially.
Graph-based methods convert the code changes from a code commit into a graph structure, incorporating control flow~\cite{cfg} or data flow dependencies~\cite{Dataflow/dfg}, and then use Graph Neural Networks (GNNs)~\cite{ggnn} or serialize the graph structure to identify security patches. For example, GraphSPD~\cite{DBLP:conf/sp/WangWSJWL23/GraphSPD} proposes a PatchCPG and employs a Graph Convolutional Network (GCN)~\cite{kipf2016gcn} to detect security patches.

\begin{figure*}[t]
	\centering
	\includegraphics[width=1.0\textwidth]{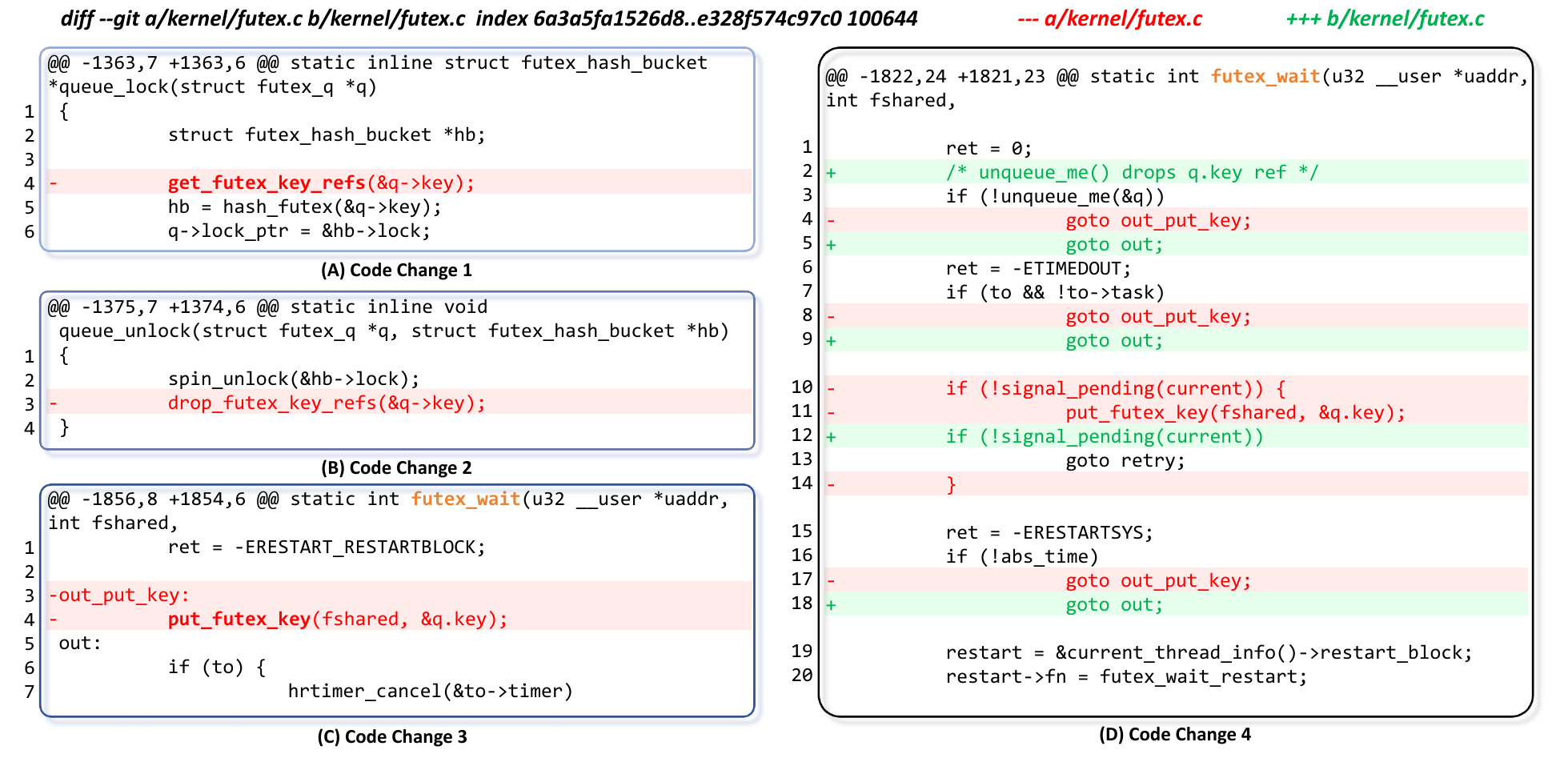}
    \caption{The part of security patch that fixes
    a buffer overflow vulnerability (i.e., CWE-119~\cite{CWE-119}). 
    The red and green lines represent the before-fixed code (pre-patch) and after-fixed (post-patch), respectively.}
\label{fig:codeexample}
\end{figure*}

However, existing methods face the following challenges: \textbf{(1) Lack of consideration of the comprehensive contexts at the repository level.}
The previous methods~\cite{DBLP:conf/sp/WangWSJWL23/GraphSPD, DBLP:conf/milcom/WangWFSJBG21/patchrnn} primarily focus on the information within the security patches,  overlooking the complex dependencies in the repository. However,
patches typically have complex dependencies in the repository, such as the call function to invoke functions in the patches, which are necessary for identifying a security patch.
For instance, as shown in Fig.~\ref{fig:codeexample}, the patch addresses a buffer overflow vulnerability identified as CVE-2014-0205~\cite{CVE-2014-0205}.
The function \texttt{futex\_wait()} (shaded in orange in Fig.~\ref{fig:codeexample}(C) and (D)) fails to properly manage a specific reference count during requeue operations, 
which may lead to trigger the use-after-free or system crash via the \texttt{queue\_lock()} and \texttt{queue\_unlock()} functions (in Fig.~\ref{fig:codeexample}(A) and (B), respectively). 
Moreover, the provided patch does not include repository-level dependency about the \texttt{get\_futex\_key\_refs} (Line 4 in Fig.~\ref{fig:codeexample}(A)) and \texttt{put\_futex\_key} (Line 3 in Fig.~\ref{fig:codeexample}(B), respectively). Therefore, we cannot determine if this patch is a valid security patch due to insufficient contextual information. 
\textbf{(2) Hard to learn the patch representation
due to the complex relationships among multiple code changes.
} 
Security patches generally encompass multiple functions and files,  increasing the difficulty in learning the
representations. For instance, the given commit involves five functions and seven code changes (Fig.~\ref{fig:codeexample} shows three functions and four code changes). These code changes do not represent a sequential relationship (i.e., they are not linearly related, such as Fig.~\ref{fig:codeexample}(C) and Fig.~\ref{fig:codeexample}(D) simultaneously modifying \texttt{futex\_wait()} function, shaded in orange). This complexity can limit the existing models' capability
to learn representations among multiple code changes.

\textbf{Our work.} To alleviate the above challenges, we propose a \textbf{\textit{Repo}}sitory-level \textbf{\textit{S}}ecurity \textbf{\textit{P}}atch \textbf{\textit{D}}etection framework named \textbf{\textit{\tool}}, which comprises three key components:  
1) a novel graph structure, called \graph, aims at extracting comprehensive contexts at the repository level by merging pre-patch and post-patch source code and retaining code changes semantics within the patches;
2) a structure-aware patch representation, which fuses the graph-based and sequence-based representations,
aiming at comprehending the relationship among multiple code changes from the structure and semantics perspective, respectively;
3) progressive learning, which aims at facilitating the model in balancing structural and semantic information. 
Additionally, we extend the SPI-DB~\cite{DBLP:journals/tosem/ZhouSWLL22/SPIDB} and PatchDB~\cite{DBLP:conf/dsn/WangWF0J21/patchdb} datasets to incorporate a total of 20,238 and 28,781 versions of repositories, respectively, denoted as SPI-DB* and PatchDB*. 

To evaluate \tool, we compare \tool with five existing security patch detection baselines and five static vulnerability detection methods.
The experimental results show that \tool outperforms the state-of-the-art security patch detection approaches, with improvements of 11.90\%, and 3.10\% in terms of accuracy and F1 score, respectively. 
Furthermore, \tool detects 151 security patches with an accuracy of 78.65\%, which achieves a substantial improvement of 21.36\% over the static-analysis-based baselines.
These results demonstrate the effectiveness of \tool in identifying security patches.

\textbf{Contributions.} The major contributions of this paper are summarized as follows:
\begin{enumerate}
\item To the best of our knowledge, we are the first to propose the repository-level patch CPG, which integrates the repository-level information and retains code change semantics within the security patches. 

\item We propose \tool, a repository-level security patch detection framework
for capturing patch
patterns from both structure and semantics perspectives
by fusing the graph-based and sequence-based
representations.

\item We curate the SPI-DB* and PatchDB* datasets, and perform an extensive evaluation. The results demonstrate the effectiveness of \tool in security patch detection. 

\end{enumerate}

\section{Background}\label{sec:back} 
\begin{figure*}[t]
	\centering
	\includegraphics[width=0.99\textwidth]{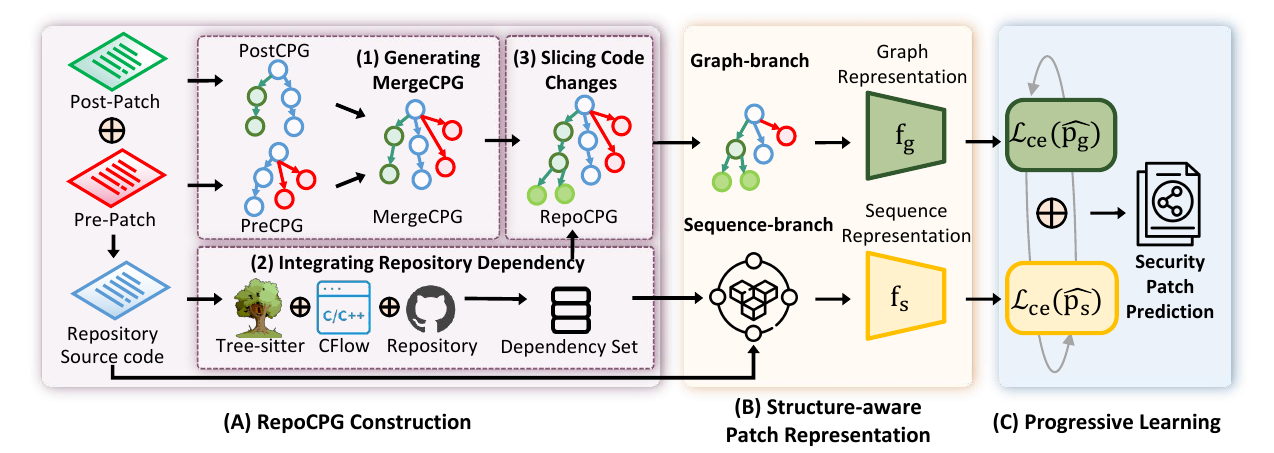}
    \caption{The overview of \tool.} 
\label{architecture}
\end{figure*}

\subsection{Code Property Graph}
Code Property Graphs (CPGs)~\cite{cpg2014} are widely used in software vulnerability-related tasks~\cite{DBLP:conf/icait/WangZWXH18, devign, reveal}. They merge Abstract Syntax Trees (ASTs)~\cite{DBLP:conf/icse/ZhangWZ0WL19/ASTNN}, Control Flow Graphs (CFGs)~\cite{cfg}, and Program Dependence Graphs (PDGs)~\cite{Dataflow/dfg} to obtain a joint graph. The PDG~\cite{IVDETECT} is divided into the Control Dependency Graph (CDG) and Data Dependency Graph (DDG), which represent control and data dependencies, respectively. For example, AMPLE~\cite{DBLP:conf/icse/WenCGZZL23} explored graph simplification on CPGs and used multi-edge-based GNNs for
vulnerability detection. GraphSPD~\cite{DBLP:conf/sp/WangWSJWL23/GraphSPD} introduced PatchGPG, applying CPGs to software security patch detection. 

Although CPG encompasses comprehensive structural and semantic information about the source code, the current approaches are limited in their ability to extract repository-level dependencies, focusing primarily on files corresponding to patches. In this paper, we extend CPGs to the repository level, incorporating more semantic and structural information to comprehend multiple code changes. 

\subsection{Repository Context in Code-related Tasks}

Incorporating repository-level context for code-related tasks has been a significant challenge. These tasks introduce numerous reasoning challenges based on real software engineering subtasks~\cite{DBLP:journals/pacmse/BairiSKCIPRAS24, DBLP:journals/corr/abs-2402-14323, DBLP:journals/corr/abs-2402-16667, DBLP:journals/corr/abs-2403-06095}, such as identifying relevant code, recognizing cross-file dependencies, and understanding repository-specific symbols and conventions.
For example, Liu et al. present RepoBench~\cite{DBLP:journals/corr/abs-2306-03091/repobench}, a benchmark specifically designed for evaluating repository-level code completion. Similarly, Wen et al. propose VulEval~\cite{DBLP:journals/corr/abs-2404-15596/vulEval}, which integrates repository-level context for vulnerability detection.

However, previous studies on identifying security patches have still relied on file- and patch-level contexts. This limitation presents challenges in real-world software production, as it is crucial for developers to be aware of other files within the repository during programming. In this paper, we further enrich
the widely-used SPI-DB and PatchDB datasets at the repository level, which highlights the future directions at the repository level security patch detection.
\section{Proposed Framework}\label{sec:med}
We provide an overview of \tool workflow in Fig.~\ref{architecture}. \tool mainly consists of three components: 
(A) \graph construction, (B) structure-aware patch representation, and (C) progressive learning.

\subsection{\graph Construction}

\setlength{\algomargin}{3em} 
\begin{algorithm}[!tpb]

    \SetAlgoLined
    
    \footnotesize
    \SetKwInOut{Input}{Input}
    \SetKwInOut{Output}{Output}

    \SetKwInOut{Initialize}{Initialize}
    \SetKwFunction{RepoCPG Construction}{RepoCPG Construction}
    \SetKwProg{Fn}{Function}{:}{}
    \Input{Pre\_code: $Code\_pre$, Post\_code: $Code\_post$, Repo\_Function: $Repo\_func$ }
    \Output{RepoCPG, $RepoCPG$}
    \Fn{RepoCPG Construction}{
    // Generating the Pre-Patch and Post-Patch CPGs
    
    $Pre\_CPG(N\_pre, E\_pre) \leftarrow Code\_pre$ 
    
    $Post\_CPG(N\_post, E\_post) \leftarrow Code\_post$

    // Fuse the Pre\_CPG and Post\_CPG to obtain the MergeCPG.
    
    \For{$v \leftarrow N\_pre, N\_post, E\_pre, E\_post$}
    {
    \eIf{$v \in V\_pre \ \textbf{and} \ v \in V\_post$}{
        $v.\text{type} \gets \text{fuse}$\;
    }{
        \If{$v \in V\_pre$}{
            $v.\text{type} \gets \text{pre}$\;
        }
        
            \If{$v \in V\_post$}{
                $v.\text{type} \gets \text{post}$\;
            }
        
    }

    }
    $Merge\_CPG \leftarrow  Construct\_MergeCPG(Pre\_CPG, Post\_CPG)$

    // Integrating the repository-level dependency.
    
    \If{$node$.type $\in$ Call\_graph}
        {
        // Adopt static tool to extract $Function\_name$ in the repository

        \If{$Function\_name$ $\in$ $Repo\_func$}
        {
            $Call\_func$ $\leftarrow$ $Repo\_func[i]$
            
            // Construct the $Call\_CPG$  of $Call\_func$ 
            
            $Call\_CPG$ $\leftarrow$ $Construct\_Call\_CPG(Call\_func)$ 
            
            // Find the $root\_node$ of $Call\_CPG$
            
            Add Edge ($node, root\_node$) and mark the $root\_node$ as $R$

            // Update Merge\_CPG
            
            $RepoCPG\leftarrow$ Update($Merge\_CPG, Call\_CPG, R$)
            
        }

        }
        
    // Slicing code changes in repository-level.
    
    \If{$R$$\in$ Code Change}
        {
        
        // Deleted-based Repository Slicing

        $RepoCPG\_Deleted$ $\leftarrow$ Delete\_Slice($RepoCPG$)

        // Added-based Repository Slicing
        
        $RepoCPG\_Added$ $\leftarrow$  Add\_Slice($RepoCPG$)
        }
    $RepoCPG$ $\leftarrow$ $RepoCPG\_Deleted \cup RepoCPG\_Added  $
    
    \Return {$RepoCPG$}
    }

\caption{\graph Construction}
\label{algorithm1}

\end{algorithm}

The purpose of the \textit{Repository-level CPG (\graph)} construction is to extract comprehensive contexts at the repository level and retain code change semantics within the patches, which consists of the following three steps: 

\begin{figure}[t]
	\centering
	\includegraphics[width=0.50\textwidth]{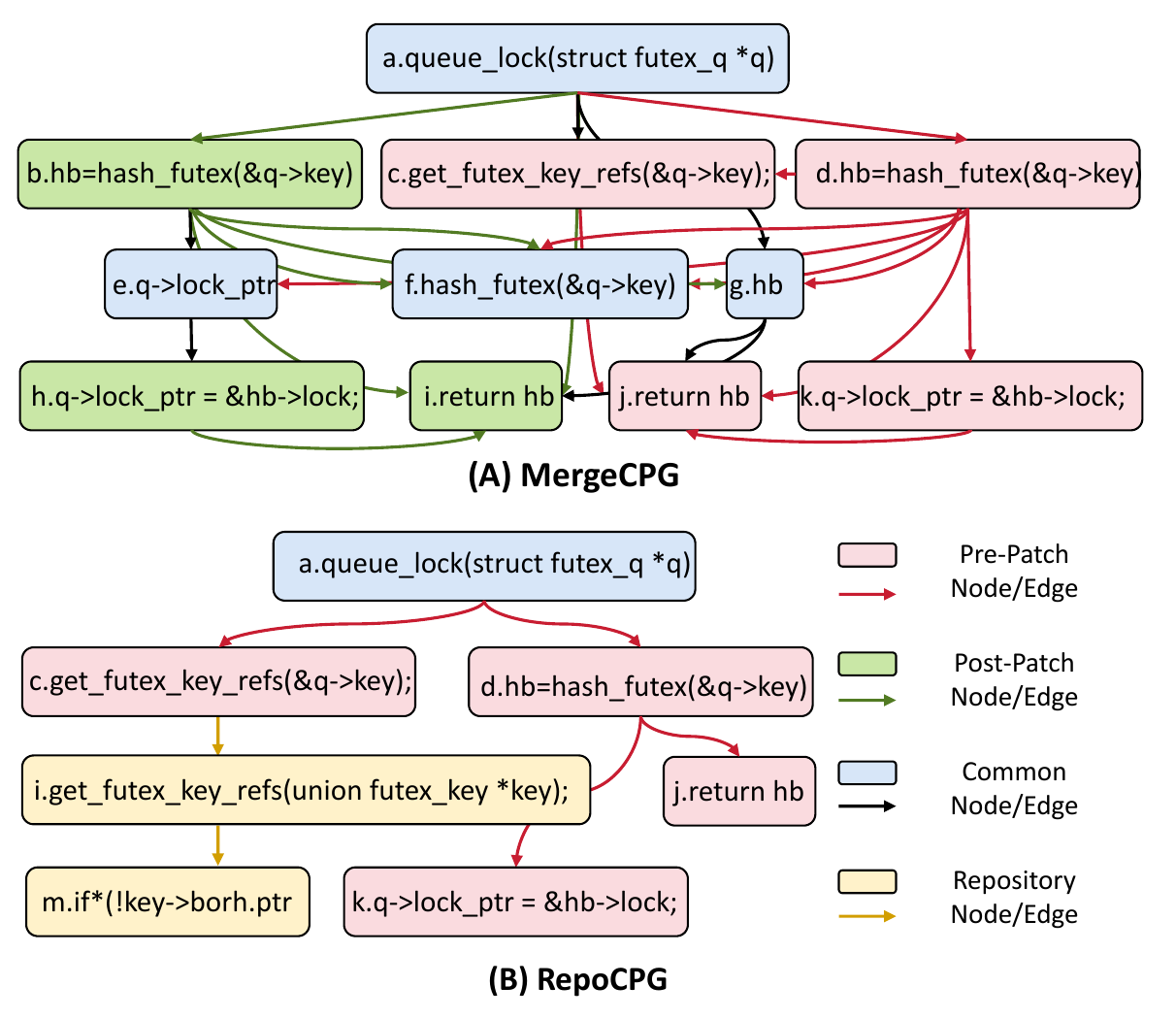}
    \caption{An example of \FuseCPG and \graph construction for the code change 1 in Fig.~\ref{fig:codeexample} (A). Due to the large size of the \graph, we ignore some nodes and edges here.}
\label{repoexample}
\end{figure}

\textit{(1) Generating \FuseCPG:}
It aims to establish connections before (pre-patch) and after (post-patch) the patch. 
However, when multiple code changes occur within a single file, they often lack structural connection to each other due to the patch only containing three lines of code within the code changes~\cite{DBLP:conf/sp/WangWSJWL23/GraphSPD}. 
Therefore, we initially employ the \textit{commit-id} in conjunction with the \textit{git reset} command to precisely revert to the specific versions of the repository, thereby retrieving the entire files both pre-patch and post-patch, rather than merely using the code changes. As shown in the
Algorithm~\ref{algorithm1} (Lines 2-4), we generate the CPG for each version of the file to construct pre-patch and post-patch CPGs, respectively.

Subsequently, we integrate the pre-patch and post-patch CPGs to establish a unified graph, designated as \FuseCPG (Lines 5-18). During the integration process, we retain \textit{common nodes} that are common across both the pre-patch and post-patch versions, such as nodes \textit{a}, \textit{e}, \textit{f}, and \textit{g}, as shown in Fig.~\ref{repoexample} (A). Nodes that exist exclusively either before or after the patch, termed \textit{changed nodes}, are also merged; for example, node \textit{c} in Fig.~\ref{repoexample} (A) appears only in the pre-patch version. We then identify and label nodes that are connected to the \textit{changed nodes}. For instance, nodes \textit{d}, \textit{j}, and \textit{k}, which connect with the \textit{changed node c}, are labeled as \textit{pre-patch nodes}. Correspondingly, nodes \textit{b}, \textit{h}, and \textit{i}, which exist in the post-patch version, are labeled as \textit{post-patch nodes}. By merging the \textit{pre-patch nodes}, \textit{post-patch nodes}, and \textit{common nodes}, we construct the \FuseCPG, which effectively preserves the semantics of the original source code and builds
the relationship between the pre-patch and post-patch. 

\textit{(2) Integrating the Repository-level Dependency:}
We then integrate the repository-level dependency (Lines 19-31 in Algorithm~\ref{algorithm1}) to obtain more contextual information.
We utilize the \textit{commit-id} to collect the corresponding specific versions of the repository.
We employ the Tree-sitter~\cite{Tree-sitter} to traverse the repository and extract the call graph~\cite{DBLP:journals/ese/KeshaniGP24/call} (Line 21). We annotate all function-level call dependencies (i.e., identified as related dependencies) and construct the repository set $Repo_{func}$ for each patch. Subsequently, we select the \textit{change code} in \FuseCPG and utilize the Cflow~\cite{cflow}, to extract dependency elements and further construct the \graph (Lines 23-29).
Specifically, since each node in the \FuseCPG contains detailed node information, we identify potential call relationships $Call_{func}$, mark the target nodes, and retrieves within the repository set $Repo_{func}$. For example, as shown in Fig.~\ref{repoexample} (B) if a dependency is identified in repository (such as node \textit{c}), we build an edge from the target node \textit{c} to the root node \textit{i} of the function. We then generate the \graph and mark node types (node \textit{i} and \textit{m}) according to its affiliation with the node version (\textit{pre-patch nodes}, \textit{post-patch nodes}, or \textit{common nodes}).
We integrate \FuseCPG with repository-level dependencies to construct \graph.

\textit{(3) Slicing Code Changes in Repository-level:}
\tool further generates a slicing \graph from the original \graph, which includes
two steps: Deleted-based repository slicing and Added-based repository slicing.
\textbf{Deleted-based repository slicing} targets the statements deleted in the pre-patch CPG. For instance, Line 4 in Fig.~\ref{fig:codeexample} (A) is a deleted statement in the pre-patch, which calls the \texttt{get\_futex\_key\_refs} function. We retain the context retrieved from the repository and incorporate it with the pre-patch statement into \graph (as discussed in Section~\ref{sec:med}A step (2)). Other context lines that have no dependency on the deleted statement are excluded.
\textbf{Added-based repository slicing} focuses on the statements added in the post-patch CPG. For example, in Fig.~\ref{fig:codeexample} (D), Line 10 is an added statement in the post-patch, which calls the \texttt{signal\_pending} function. We retain the context and incorporate it with the post-patch statement into \tool. 
Both deleted-based and added-based repository slicing are conducted according to the types of edges within the CDG and DDG, where each node represents a statement. All retained nodes in \graph are directly dependent on the code changes in the patch. Fig.~\ref{repoexample} (B) shows
an example of \graph after code change slicing
at the repository level.

\subsection{Structure-aware Patch Representation}
In this component, we elaborate on the proposed structure-aware patch representation, which involves the graph branch for capturing the structural information of \graph, and combines the sequence branch for further enhancing the security patch representations.

\textit{(1) Graph branch:} The graph branch learns vulnerability patterns from semantic and structural information. The semantic information is exhibited by each node embedding in the \graph, while the structural information is achieved by the graph structure by diverse relationships between node and edges. The content of each node in \graph can be a statement node in CDG/DDG or token node in AST. We use  UniXcoder~\cite{DBLP:conf/acl/GuoLDW0022/unixcoder} to initialize the representations of each node in the \graph. We generate the node vector $h_i^{(0)}$ for each node $i$ in the \graph, which is calculated as follows:

\begin{equation}
\label{cls}
h_i^{(0)} = H^{0}(N_i) +  H^{L}(N_i)
\end{equation}
where $H^{0}$ and $H^{L}$ denote the first and last layer embedding of UniXCoder~\cite{DBLP:conf/acl/GuoLDW0022/unixcoder}, respectively.

We then use the Graph Attention Networks (GAT)~\cite{DBLP:journals/corr/abs-1710-10903/gat}, which aims at capturing the local structural information of the \graph.
Specifically, two types of edge roles and four-bit vectors are defined: version (\textit{pre-patch}, \textit{post-patch}, or \textit{common}), and functional (\textit{CFG}, \textit{DDG}, or \textit{AST}). It is noteworthy that repository-level edges are not listed separately; instead, they encompass both version information and functional relationships. 
Owing to the distinct roles, it is impractical to apply a uniform set of weights across the entire model for learning the graph representation. Consequently, we construct four subgraphs (i.e., four GAT layers) to cater to four-bit vectors. Within this framework, a GAT layer computes a new set of node embedding by leveraging the input node features along with the attention coefficients that have been learned. For each subgraph, the normalized attention coefficients between nodes $i$ and $j$ are calculated using the following formula:
\begin{equation}
\alpha^{(l)}_{ij}[k] = \text{softmax}\left( \text{LeakyReLU}\left(a^{(l)} \left[ W h^{(l-1)}_i \parallel W h^{(l-1)}_j \right]\right) \right),
\end{equation}
where $a^{(l)}$ and $W$ denote the learnable vector and weight matrix, respectively. $\parallel$, \text{softmax}, and \text{LeakyReLU} denote the concatenation operation, softmax function, and activation function, respectively. $k$ is the index of the subgraph.
The node embedding $h^{(l)}[k]$ of subgraph $k$ in layer $l$ is computed as follows:
\begin{equation}
h^{(l)}_i[k] = \sigma \left( \sum_{j \in N(i)} \alpha^{(l)}_{ij}[k] W[k] h^{(l-1)}_j \right),
\end{equation}
where $N(i)$ and $\sigma$ denote the set of neighboring nodes of node $i$ and the activation function, respectively. The different roles of edges and the structure-level information in the \graph will be aggregated to the whole graph feature $h^{(l)}_i$, which can be formulated as:
\begin{equation}
h^{(l)}_i = \parallel_{k=1}^{K} h^{(l)}_i[k]
\end{equation}
where $K$ represents the total number of subgraphs. Since the graph feature $h^{(l)}_i$ only provides individual attention for each subgraph, we further employ another GAT layer to learn the structural information from the whole \graph. Finally, we calculate the representation $h^{(l)}_i$ through a pooling layer to obtain the graph-branch representation $f_{g}$. 

\textit{(2) Sequence branch:} We also integrate a sequence branch specifically designed to analyze the code changes within a patch. A typical patch includes several lines of contextual code, but it also often encompasses a substantial amount of extraneous data, such as line numbers and diff markers (e.g., "@@ string1 @@"). Such information can potentially mislead the model during the learning process. To mitigate this issue, the sequence branch is configured to selectively retain only the lines of code changes, along with the version information (either pre-patch or post-patch), to enhance semantic learning. We exclude non-critical elements such as index lines, filenames, and location indicators.
Specifically, we fine-tune~\cite{DBLP:journals/tmi/TajbakhshSGHKGL16/finetune} the UniXcoder~\cite{DBLP:conf/acl/GuoLDW0022/unixcoder} and obtain the sequential representation $f_{s}$ for the sequence branch.

\subsection{Progressive Learning}

As shown in Fig.~\ref{architecture}(C), we introduce progressive learning that systematically alternates focus between the graph and sequence branches by modulating the respective model weights. This approach is necessitated by the inherent differences in modalities and input characteristic between
the graph and sequence branches. Due to these differences, the learning rates and shared parameters between the branches vary substantially, posing challenges for capturing patch-related
patterns. Therefore, we propose a method to progressively learn the graph branch and sequence branch, by selectively freezing the model weights of branches during the training process.

More concretely, the feature vectors $f_{g}$ and $f_{s}$ will be sent into the classifiers $W_{g}$ and $W_{s}$ respectively and the outputs will be integrated together. The predicted output is formulated as:
\begin{equation}
p = \frac{W^\top_{g}f_{g} + W^\top_{s}f_{s}}{2}
\end{equation}

Then, progressive learning is to initially train the sequence model, leveraging the domain knowledge provided by the pre-trained model. Subsequently, the focus shifts to the more complex graph structure, ensuring comprehensive learning across different data representations. The progressive learning loss of \tool is illustrated as:

\begin{equation}
\mathcal{L} = T\mathcal{L}_{ce}(\hat{p_{g}}, y) + (1 - T)\mathcal{L}_{ce}(\hat{p_{s}}, y), T = 
\begin{cases} 
0 & \text{if } E \geq \frac{E_{max}}{2} \\
1 & \text{if } E < \frac{E_{max}}{2} 
\end{cases}    
\end{equation}
where $\mathcal{L}_{ce}$ and $y$ denote the cross-entropy loss function and the label of data, respectively. $\hat{p_{g}} = W^\top_{g}f_{g}$ and $\hat{p_{g}} = W^\top_{s}f_{s}$ are the predicted output of graph and sequence branch, respectively. $T$ denotes the shifting parameter and $E$ denotes the current training epoch.

\section{Experimental setup}\label{sec:setup}
\subsection{Research Questions}

In this section, we evaluate the effectiveness of \tool by comparing it with the state-of-the-art security patch detection approaches and focus on the following four Research Questions (RQs):

\begin{enumerate}[label=\bfseries RQ\arabic*:,leftmargin=.5in]
    \item How effective is \tool compared with existing security patch detection approaches?
    \item How effective is \tool in security patch detection compared with the 
    static analysis-based approaches?
     \item How effective is \tool over patches with different vulnerability types?
    \item What is the influence of different components of \tool on the performance for identifying security patches? 
\end{enumerate}

\subsection{Datasets}
\begin{table}[t]
\centering
\setlength{\tabcolsep}{2mm}
\renewcommand{\arraystretch}{1.3}

\caption{Statistics of the SPI-DB* and PatchDB* datasets.}
\resizebox{0.47\textwidth}{!}{
\begin{tabular}{ccccc}

\toprule
\rowcolor[HTML]{DEDEDE}
\textbf{Dataset} & \textbf{Set} & \textbf{\#Patch} &  \textbf{\#File} & \textbf{\#Version of Repo} \\
\midrule
\multirow{4}{*}{SPI-DB*}  & Train & 16,454 & 64,580 & 16,296 \\
 & Valid & 2,028  & 7,994 & 2,024 \\
 & Test & 2,000  & 7,649 & 1,996 \\
\cline{2-5}
 & All & 20,482 & 80,223  & 20,238 \\
\midrule
\multirow{4}{*}{PatchDB*} &Train  & 23,229 & 82,913 & 23,047  \\
                   & Valid & 2,907  & 10,349 & 2,903 \\
                   & Test & 2,906  & 9,996 & 2,900 \\
                   \cline{2-5}
                   & All & 29,042 & 103,258 & 28,781 \\
\bottomrule
\end{tabular}
}
\label{dataset}
\end{table}

\subsubsection{Data Source}

To address the proposed RQs, we select two widely-used datasets as the raw data: SPI-DB~\cite{DBLP:journals/tosem/ZhouSWLL22/SPIDB} and PatchDB~\cite{DBLP:conf/dsn/WangWF0J21/patchdb}. Specifically, SPI-DB collects patches from two major C/C++ datasets, FFMPeg and Qemu, encompassing 25k patches, of which 10k have been classified as security-related. PatchDB compiles data from 348 open-source repositories, containing over 36k code snippets, approximately 12k identified as security patches. 

\subsubsection{Data Process}
To evaluate the security patches at the repository level, we further collect the versions of the repository source code and extract the dependency at the repository level via three steps: 
(1) 
We initially select repositories from which complete source code and commit logs can be retrieved via GitHub. This process resulted in the selection of 20,482 patches from SPI-DB and 29,042 from PatchDB, as detailed in Table~\ref{dataset}.
(2) 
For each identified patch, we use the corresponding commit ID to collect specific code versions at the repository level. This effort has led to the collection of 20,238 versions from repositories (i.e., \#Version of Repo in Table~\ref{dataset}) in SPI-DB and 28,781 from PatchDB.
(3) Finally, we traverse the entire repository by Tree-sitter~\cite{Tree-sitter} tool to parse the function-level dependencies. Subsequently, we employ the CFlow~\cite{cflow} to further extract 
dependency elements. As detailed in Table~\ref{dataset}, we extract the repository dependencies from 80,223 and 103,258 files to construct the SPI-DB* and PatchDB*, respectively.

\subsubsection{Data Split}
Following the previous work~\cite{DBLP:conf/sp/WangWSJWL23/GraphSPD}, we split the datasets into disjoint training, validation, and test sets in a ratio of 8:1:1, as shown in the Table~\ref{dataset}. We use the training set to train the models, use the validation set for selecting best-performance models, and evaluate the performance in the test set.

\subsection{Baselines}
\subsubsection{Comparison on Security Patch Detection Approaches}

To address RQ1, we evaluate the effectiveness of \tool\ by comparing the following six security patch detection approaches. We categorize these baselines into three groups:
\textbf{Supervised-based methods:} We utilize PatchRNN and GraphSPD for this category. PatchRNN~\cite{DBLP:conf/milcom/WangWFSJBG21/patchrnn} employs an RNN-based model that processes only source code as input, while GraphSPD~\cite{DBLP:conf/sp/WangWSJWL23/GraphSPD} introduces PatchCPG and leverages a GNN to learn structural information. These methods are widely recognized and frequently adopted as baselines in recent studies.
\textbf{Pretrained model-based methods:} We select three prominent pre-trained models:
CodeBERT~\cite{DBLP:conf/emnlp/FengGTDFGS0LJZ20/codebert}, CodeT5~\cite{DBLP:conf/emnlp/0034WJH21/CodeT5}, and UniXcoder~\cite{DBLP:conf/acl/GuoLDW0022/unixcoder}. These models use code changes (i.e., patches) as input and are further fine-tuned for the downstream task of security patch detection.
\textbf{LLM-based methods:} Due to resource constraints, we construct the prompt and utilize Llama3-70b~\cite{llama3} to assess the performance of LLMs in security patch detection.

\subsubsection{Comparison on Static Analysis Approaches}
In RQ2, beyond directly identifying security patches, a common approach involves utilizing static analysis tools to detect security patches~\cite{DBLP:conf/sp/WangWSJWL23/GraphSPD}. This method entails detecting vulnerabilities in pre-patch code snippets and verifying their absence in post-patch code snippets. In this paper, we select five widely used baselines: Cppcheck~\cite{Cppcheck}, RATS~\cite{RATs}, Semgrep~\cite{Semgrep}, Flawfinder~\cite{FLAWFINDER}, and VUDDY~\cite{DBLP:conf/sp/KimWLO17/vuddy}. 
These methods employ predefined rules and patterns to identify potential vulnerabilities in source code at the repository level. 
Therefore, they can detect vulnerabilities in the vulnerable version (i.e., pre-patch) and should not identify such vulnerabilities if they have been successfully patched (i.e., post-patch).


\subsection{Evaluation Metrics}
We choose the following three metrics to evaluate the performance of \tool.
\begin{itemize}
    \item \textbf{Accuracy:} $Accuracy=\frac{TP+TN}{TP+TN+FN+FP}$. It reflects the percentage of the samples which are correctly classified among samples. $TP$ and $TN$ denote the counts of true positive and true negative samples, respectively. $TP+TN+FN+FP$ represents the total number of samples.


    
    \item \textbf{F1 score:} $F1=\frac{2\times Precision\times Recall}{Precision+Recall}$. F1 is the harmonic mean of precision and recall.
     
    \item \textbf{False Positive Rate (FPR):} $FPR=\frac{FP}{FP+TN}$. It measures the proportion of negative samples that are erroneously classified as positive. 
     
     
\end{itemize}

\subsection{Implementation Details}

For all the baselines except Llama3-70b, we directly use the publicly available source code and hyper-parameters released by the paper.   
For Llama3-70b, we use the Ollama~\cite{ollama} framework and evaluate them on our server. 

To ensure the fairness of all experiments, we consistently apply the same data across all approaches. We utilize Joern~\cite{Joern} to generate distinct CPGs~\cite{cpg2014} for both pre-patch and post-patch code. Additionally, Scala is employed to construct the \FuseCPG. Then, we use the Tree-sitter~\cite{Tree-sitter} and Cflow~\cite{cflow} to analyze and construct the \graph. 
Tree-sitter is utilized to extract function-level source code by traversing the repository, thereby facilitating subsequent analysis. It is a static analysis tool, which can quickly parse and analyze code structures across repositories~\cite{DBLP:conf/issta/WenGGXL24}.
Cflow is employed to extract dependencies at the repository level, focusing on the flow of function calls~\cite{DBLP:journals/corr/abs-2404-15596/vulEval}. For the graph node embedding, we leverage the pre-trained model UniXcoder~\cite{DBLP:conf/acl/GuoLDW0022/unixcoder}, utilizing its tokenization and model weight to obtain initial node vectors. In the sequence branch, we fine-tune the UniXcoder model with a learning rate of $2 \times 10^{-5}$. 

For the graph branch, we use a learning rate of $5 \times 10^{-5}$ and set the GAT with 2 heads. The training process spans 10 epochs with a batch size of 4. 
We conduct experiments to determine the suitable hyper-parameters. Due to space limitations, the experimental results of PatchDB* are shown in GitHub. 
All experiments are conducted on a server equipped with NVIDIA GeForce RTX 3090 GPU and CUDA 11.3.
\section{Experimental Results}\label{sec:result}

\begin{table}[t]
\centering
\setlength{\tabcolsep}{3mm}
\renewcommand{\arraystretch}{1.2}

\caption{Experimental results of \tool and the security patch detection baselines on the SPIDB* and PatchDB* datasets. 
Texts in bold represent the best performance of the best methods in each metric. }
\resizebox{0.47\textwidth}{!}{
\begin{tabular}{c|c|ccccc}

\toprule
\rowcolor[HTML]{DEDEDE}
\textbf{Dataset}                            & \textbf{Method}             & \textbf{Accuracy$\uparrow$}                               & 
\textbf{F1 Score$\uparrow$}                               & \textbf{FP Rate$\downarrow$}                                                                               \\ \toprule
   & GraphSPD         & 59.40           & 59.11          & 22.08          \\
                          & PatchRNN         & 57.95          & 60.46          & 43.97          \\
                          & CodeBERT         & 65.61          & 61.40           & 32.56          \\
                          & CodeT5           & 66.62          & 61.87          & 30.31          \\
                          & UniXCoder        & 66.27          & 60.26          & 28.27          \\
                          & Llama3-70b       & 56.35          & 42.98          & 26.38          \\
\multirow{-7}{*}{SPIDB*} &\textbf{ \tool } & \textbf{74.55*}                          & \textbf{68.98}        & \textbf{14.67}               \\ \midrule
 & GraphSPD         & 71.42          & 48.67          & 16.52          \\
                          & PatchRNN         & 70.15          & 30.45          & 8.23           \\
                          & CodeBERT         & 78.27          & 65.27          & 16.56          \\
                          & CodeT5           & 80.84          & 58.52          & 9.28           \\
                          & UniXCoder        & 80.70           & 64.34          & 8.68           \\
                          & Llama3-70b       & 74.65          & 55.84          & 15.43          \\
\multirow{-7}{*}{PatchDB*}                  &\textbf{ \tool }    & \textbf{83.35*}                                         & \textbf{69.13}                                           & \textbf{6.65}                                          \\
\bottomrule

\end{tabular}
}
\label{RQ1}
\end{table}

\subsection{RQ1: \tool VS. Security Patch Detection Approaches}
To answer RQ1, we compare the three types of security patch detection approaches, including supervised-based, pre-trained model-based, and LLM-based methods.
Table~\ref{RQ1} shows the experimental results of each baseline of accuracy, F1 score, and FP rate metrics. 

\subsubsection{Overall Results} 
The experimental results presented in Table~\ref{RQ1} demonstrate that \tool consistently outperforms all baseline methods on the SPI-DB* and PatchDB* datasets across all evaluated metrics. Specifically, \tool achieves an accuracy of up to 74.55\%, and an F1 score of 68.98\% on the SPI-DB* dataset. Furthermore, \tool exhibits a higher accuracy of 83.35\% and a reduced FP rate of 6.65\% on the PatchDB* dataset. This enhanced performance in PatchDB* may be attributed to the larger training data compared to SPI-DB*.

\subsubsection{\tool VS. Supervised-based Methods}
The results summarized in Table~\ref{RQ1} indicate that \tool enhances performance metrics across both datasets when compared to all other supervised-based methods. Specifically, \tool achieves average improvements of 14.22\% in accuracy, and 20.61\% in F1 score. GraphSPD outperforms TwinRNN, primarily because GraphSPD integrates the dependency within the patch to provide structural information. Furthermore, \tool incorporates repository-level information to construct \graph, which contributes to its superior performance.
\subsubsection{\tool VS. Pre-trained Model-based Methods}
Table~\ref{RQ1} reveals that pre-trained model-based methods generally surpass those supervised-based methods. Specifically, CodeBERT, CodeT5, and UniXcoder achieve an average performance of 
66.17\% and 61.18\% in terms of the accuracy and F1 score on the SPI-DB* dataset. 
Despite these results, these methods still behave worse than \tool across all metrics. Particularly, \tool outperforms the state-of-the-art baseline, with relative improvements of 7.50\% of accuracy, and 8.70\% of F1 score on average across the SPI-DB* and PatchDB* datasets. 
The primary factor contributing to this performance gap is that pre-trained model-based methods focus on semantic information within code changes. However, they do not adequately address the lack of structural information at the repository level, which is crucial for comprehensive analysis. In contrast, \tool effectively integrates both semantic and structural information, thereby enhancing its overall performance in security patch detection. Furthermore, we also conduct the statistical significance tests between \tool and the best-performing methods, CodeT5. \tool surpasses CodeT5 in both PatchDB* and SPI-DB* at the 0.05 significance level, with p-values of 1.15E-2 and 5.46E-7, respectively. These results show that the RepoSPD significantly outperforms other baselines.

\subsubsection{\tool VS. LLM-based Methods}
Due to resource constraints, we only select Llama3-70b for evaluating the performance of LLMs in security patch detection. 
As indicated in Table~\ref{RQ1}, \tool surpasses the performance of Llama3-70b in all datasets and metrics, despite the generally robust capabilities of LLM-based methods across various domains.
There are two primary reasons that may lead
Llama3-70b to failing
to identify security patches. Firstly, the single prompt may result in a lack of domain-specific knowledge in security patch detection, thereby rendering the detection of complex vulnerability types particularly challenging.
Secondly, the model struggles with the lack of repository-level dependencies between code changes at the repository level, which often leads to an inability to make definitive judgments.

\begin{tcolorbox}
[width=\linewidth-2pt,boxrule=3pt,top=3pt, bottom=1pt, left=3pt,right=3pt, colback=gray!20,colframe=gray!25]
 \textbf{Answer to RQ1:} In comparison with the security patch detection approaches, \tool achieves the best performance across all evaluated performance metrics, with improvements of 7.50\% of accuracy, and 8.70\% of F1 score on average across the SPI-DB* and PatchDB* datasets.
 \end{tcolorbox}

\subsection{RQ2: \tool VS. Static Analysis Approaches}
\begin{table}[t]
\centering
\setlength{\tabcolsep}{1.5mm}
\renewcommand{\arraystretch}{1.2}

\caption{The experimental results of \tool and the static vulnerability detection baselines on the PatchDB* dataset. 
Texts in bold represent the best performance of the best methods in each metric. }
\resizebox{0.5\textwidth}{!}{
\begin{tabular}{l|cc|cc}

\toprule
\rowcolor[HTML]{DEDEDE}
\textbf{Method} & \multicolumn{1}{l}{\#\textbf{Vul\textsubscript{pre-patch}}} & \#\textbf{Vul\textsubscript{post-patch}} & \#\textbf{Security Patch$\uparrow$} & \multicolumn{1}{l}{\textbf{Accuarcy(\%)$\uparrow$}} \\
\midrule
Cppcheck           & 31                                           & 31                        & 0                       & 0.00                                 \\
RATS               & 109                                          & 110                       & 0                       & 0.00                                 \\
Semgrep            & 16                                           & 20                        & 0                       & 0.00                                 \\
Flawfinder         & 148                                          & 150                       & 5                      & 2.60                                \\
VUDDY              & 131                                          & 59                        & 110                     & 57.29                               \\
\tool                &       -                                       &    -                       & \textbf{151}                     & \textbf{78.65}                                     
\\
\bottomrule

\end{tabular}
}
\label{RQ2}
\end{table}

To evaluate the effectiveness of \tool in identifying security patches, we also compare
with widely-used static analysis-based approaches. 
In this paper, we select five static analysis tools and utilize a dataset comprising 192 security patches.
Specifically, to maintain consistency with GraphSPD~\cite{DBLP:conf/sp/WangWSJWL23/GraphSPD}, the criteria of the selection include: (1) The patches without associated CVEs are removed since they are hard to be verified. (2) The non-security patches are excluded from the dataset. (3) The patches are selected only from the test set to prevent data leakage between the training and valid sets.
Based on the selection criteria, the Patch-DB* comprises only 192 patches.

Table~\ref{RQ2} presents the number of vulnerabilities detected by static analysis-based techniques in both pre-patch and post-patch code. As demonstrated in Table~\ref{RQ2}, the results show that \tool surpasses all baselines, detecting an additional 41 security patches and achieving a 21.36\% improvement in terms of accuracy. 
The static analysis tools such as Cppcheck, RATS, and Semgrep fail
to detect any security patches, indicating their
potential limitation in 
accurately identifying
identify security patches. 
For instance, Cppcheck identifies 31 vulnerabilities in the pre-patch versions of the code. However, it also identifies the same vulnerabilities in the post-patch versions, indicating that it fails to recognize any of the applied security patches.

Among the static analysis-based approaches, VUDDY exhibits superior detection performance, identifying 131 vulnerabilities in pre-patch code and 59 in post-patch code. Among the vulnerabilities detected in the post-patch code, 38 are not associated with the same patches as the 131 vulnerabilities identified in the pre-patch code. This discrepancy highlights the challenges in accurately classifying patches, with VUDDY correctly identifying 110 patches as secure patches. 

In summary,
despite the widespread use of existing static analysis-based methods in practice, they exhibit a notable deficiency in detecting security patches. This limitation underscores the value of \tool, which demonstrates a robust capability to identify security patches in the repository-level, thereby enhancing its practical utility. 

\begin{tcolorbox}
[width=\linewidth-2pt,boxrule=3pt,top=3pt, bottom=1pt, left=3pt,right=3pt, colback=gray!20,colframe=gray!25]
 \textbf{Answer to RQ2:} In comparison with the static analysis-based approaches, \tool surpasses all baselines, detecting an additional 41 security patches and achieving a 21.36\% improvement in terms of accuracy. 
 \end{tcolorbox}
\subsection{RQ3: Effectiveness of Different Types of Patches in \tool}

To evaluate the effectiveness of \tool in identifying different types of security patches, we utilize the same dataset as in RQ2. This dataset encompasses 28 types of Common Weakness Enumerations (CWE)~\cite{CWE} across 74 projects. The proportion and type of vulnerability are presented in Table~\ref{RQ3}. It is important to note that this dataset exclusively contains security patches and does not include non-security patches.

Overall, we observe that \tool is effective across all ten types of vulnerabilities analyzed, achieving an average accuracy of 81.11\%. 
We can observe the following findings:
(1) The \tool shows excellent performance at identifying vulnerabilities associated with a higher number of security patches, such as Buffer Overflow, Resource Leakage, and Numeric Error, with accuracy performance of 87.88\%, 82.14\%, and 91.67\%, respectively. In addition, it also reveals that \tool with a relatively low number of security patches can perform well in some cases, such as untrusted data and race conditions.
(2) The security patches that are frequently misclassified typically pertain to vulnerabilities stemming from inadequate verification processes, such as improper input validation and improper access control. These cases often do not present overt errors but rather require the understanding of multiple dependencies to identify vulnerabilities. 

\begin{tcolorbox}
[width=\linewidth-2pt,boxrule=3pt,top=3pt, bottom=1pt, left=3pt,right=3pt, colback=gray!20,colframe=gray!25]
 \textbf{Answer to RQ3:} \tool shows excellent performance in identifying vulnerabilities, especially with a higher number of security patches. 
 The security patches that are frequently misclassified typically pertain to vulnerabilities stemming from inadequate verification processes.  
 \end{tcolorbox}
\begin{table}[t]
\centering
\setlength{\tabcolsep}{2.5mm}
\renewcommand{\arraystretch}{1.2}

\caption{ The security patch detection performance of \tool over different vulnerability types.}
\resizebox{0.47\textwidth}{!}{
\begin{tabular}{c|c|cc}

\toprule
\rowcolor[HTML]{DEDEDE}
\textbf{Vulnerability Type}                              & \textbf{CWE-ID}                       & \textbf{Ratio(\%)} & \textbf{Acc(\%)$\uparrow$}      \\ 
\midrule
Buffer Overflow                   & 119, 125, 787                 & 34.38    & 87.88  \\
Double Free / Use After Free        & 415, 416                      & 3.65     & 85.71  \\
Injection                         & 74, 77, 94                    & 1.56     & 66.67  \\
Improper Input Validation         & 20, 22                        & 9.38     & 61.11  \\
Resource Leakage                  & 200, 399, 400                 & 14.58    & 82.14  \\
Numeric Error                     & 189, 190, 369                 & 12.50    & 91.67  \\
NULL Pointer Dereference          & 476                           & 6.77     & 69.23  \\
Improper Access Control           & 264, 284                      & 9.38     & 66.67  \\
Untrusted Data & 502                           & 0.52     & 100.00 \\
Race Condition                    & 362                           & 1.04     & 100.00 \\
Other Vulnerabilities                             & \begin{tabular}[c]{@{}c@{}}16, 17, 19, 59, \\ 254, 310, 426\end{tabular} & 6.25     & 41.67 
\\
\bottomrule
\end{tabular}
}
\label{RQ3}
\end{table}

\subsection{RQ4: Effectiveness of Different Components in \tool}
In this section, we explore the impact of different components of \tool including the \graph construction (i.e., w/o \graph), the structure-aware patch representation (i.e., w/o sequence and w/o graph), and progressive learning (i.e., w/o progressive).
The experimental results are shown in Table~\ref{RQ4}.
\subsubsection{\graph Construction}
To explore the effect of the \graph,  
we deploy one variant (i.e., w/o \graph) by only using the PatchCPG proposed by GraphSPD. 
As shown in Table~\ref{RQ4}, the \graph can improve the performance of \tool on all datasets. 
Specifically, incorporating the repository-level information to construct the \graph
leads to the average drop of 5.12\% in accuracy, and 5.88\% F1 score on two datasets. Especially on the SPI-DB* dataset, \graph boosts the performance by 9.80\% for accuracy, and 11.32\% for F1 score, respectively. 
In addition, We attribute the relatively small improvement on the PatchDB* dataset to the higher number of dependencies, which is approximately twice that of SPI-DB*. The complex inter-dependencies may hinder the model's learning process to some extent.

\subsubsection{Structure-aware Patch Representation}
To investigate the impact of the structure-aware patch representation, we construct two experimental variants for comparative analysis: (1) a variant exclusively utilizes the graph-based branch (i.e., w/o sequence), and (2) another solely employs the sequence-based branch (i.e., w/o graph), which verify the effectiveness of capturing semantic and structural information for detecting security patches, respectively.

Our findings indicate a consistent performance decline across two datasets when the variants operate independently. Specifically, the variant using only the graph branch exhibits a decrease of 8.00\%, while the variant relying solely on the sequence branch shows a decrease of 5.47\% in terms of accuracy. 
This demonstrates that the sequence branch exerts more influence on \tool.
Furthermore, the graph branch notably reduces the FP rate, with the decrease of 13.60\% and 2.03\% in the SPI-DB* and PatchDB* datasets, respectively. 
This underscores the importance of structural information to increase the performance of security patch detection.
We can achieve
that both the graph and sequence branches contribute to the overall performance of \tool, each playing a crucial role in security patch detection.

\subsubsection{Progressive Learning}

\begin{table}[t]
\centering
\setlength{\tabcolsep}{2.5mm}
\renewcommand{\arraystretch}{1.2}

\caption{The experimental results of \tool in PatchDB* and SPIDB* datasets when removing the \graph (i.e., w/o 
\graph), removing the sequence branch (i.e., w/o sequence), removing the graph branch (i.e., w/o graph) and removing the progressive learning(i.e., w/o progressive) in three metrics.}
\resizebox{0.47\textwidth}{!}{
\begin{tabular}{c|c|ccc}

\toprule
\rowcolor[HTML]{DEDEDE}
\textbf{Dataset}                            & \textbf{Variant}             & \textbf{Accuracy$\uparrow$}                               & 
\textbf{F1 score$\uparrow$}                               & \textbf{FP Rate$\downarrow$}                                                                               \\ \toprule
&{w/o \graph}                     & 64.75                            & 57.66                           & 24.82                                \\

&w/o sequence                             & 69.60                   & 62.33                           & 17.99                                \\
&w/o graph                                & 66.27                    & 60.26                           & 28.27                                \\

&{w/o progressive}                        & 72.25                    & 66.91                           & 18.45                                \\

&{w/o change order}     & 67.45&	62.65	&25.92                                         \\

\multirow{-6}{*}{SPIDB*}&\tool               & 74.55                  & 68.98                           & 14.67                               
 \\
 \midrule

&{w/o \graph}                     & 82.91                             & 68.69                           & 7.41                                 \\

&w/o sequence                             & 72.31                        & 31.49                           & 4.86                                 \\
&w/o graph                                & 80.70                           & 64.34                           & 8.68                                 \\

& w/o progressive                              & 80.94                        & 65.55                           & 9.38                                 \\

&{w/o change order}     & 81.66&	68.25	& 10.62                                         \\

\multirow{-6}{*}{PatchDB*}&\tool& 83.35                              & 69.13                           & 6.65                                 \\

\bottomrule

\end{tabular}
}
\label{RQ4}
\vspace{-12pt}
\end{table}

To understand the effect of progressive learning, we also implement two variants for comparative analysis: (1) a variant of \tool without the progressive learning component (i.e., w/o progressive), thereby requiring the model to learn weights simultaneously across all components. (2) another variant first trains the graph representation and then refines it into a sequence representation (i.e., changes order).
Specifically, the performance consistently shows an average decrease of 2.36\% in accuracy, and 2.83\% in F1 score across SPI-DB* and PatchDB* datasets without a progressive learning component. It also demonstrate that \tool performs well across two datasets, with an average improvement of 4.40\% in accuracy and 3.61\% in F1 score, while reducing the FPR by 7.61\%.
 These results underscore the different branches have different learning strategies across various branches of the \tool, with each branch learning unique discriminative representations. It allows for staged optimization and more targeted learning of different branches, which in turn enhances the overall capabilities of detecting security patches.

\begin{tcolorbox}
[width=\linewidth-2pt,boxrule=3pt,top=3pt, bottom=1pt, left=3pt,right=3pt, colback=gray!20,colframe=gray!25]
 \textbf{Answer to RQ4:} All components, including \graph construction, structure-aware patch representation, and progressive learning, enhance the performance of \tool.  
 \end{tcolorbox}
\section{Discussion}\label{sec:discussion}
\subsection{Why does \tool Work?}

We identify the advantages of \tool, which can explain its effectiveness in security patch detection.

\begin{figure*}[t]
	\centering
	\includegraphics[width=0.95\textwidth]{figures/Discussion\_cropped.pdf}
    \caption{(A) and (C) represent the security patch for resource leakage vulnerability (CVE-2016-5243~\cite{CVE-2016-5243}) and buffer overflow vulnerability (CVE-2016-9535~\cite{CVE-2016-9535}), respectively. Fig. (B) is the dependency (i.e., repository information) extracted from CVE-2016-5243 to construct \graph. }
\label{disscussion}
\vspace{-12pt}
\end{figure*}

\textit{(1) Incorporating repository-level information to help the security patch detection.} 
We propose \graph to integrate repository-level dependencies,
which enhances the performance of security patch detection.
As illustrated in Fig.~\ref{disscussion}(A), the example is drawn from a resource leakage vulnerability in the Linux kernel (i.e., CVE-2016-5243~\cite{CVE-2016-5243}). Specifically, the function \texttt{tipc\_nl\_compat\_link\_dump} in net/tipc/netlink\_compat.c fails to copy a particular string (Line 5), thereby allowing local users to access sensitive data from the kernel stack memory. This issue is addressed in Lines 6-7 through the use of \texttt{nla\_strlcp}. However, \texttt{nla\_strlcp} is not defined within the patch itself. \tool effectively incorporates \texttt{nla\_strlcp} as a dependency into \graph, and accurately identifies the security patch. 

\textit{(2) Effectively capturing both structural and sequential information among multiple code changes.} 
We propose the structure-aware patch representation to comprehend the relationships
among multiple code changes.
As illustrated in Fig.~\ref{disscussion} (C), the \texttt{\_TIFFmalloc} call at Line 5 of the pre-patch code implicitly assumes successful memory allocation, which effectively acts as an implicit assertion check. This assumption may lead to buffer overflows in release mode, particularly when handling atypical tile sizes. Furthermore, the patch includes a total of 11 code changes. 
\tool can extract the structural dependency between Line 12 and Line 18, which collaboratively changes
the function \texttt{fpDiff}.
Additionally, \tool captures the semantic information, revealing that Line 4 and Line 18 utilize identical statements to address the vulnerabilities.
Based on the structural and semantic information, \tool effectively detects the security patch.

\begin{table}[t]
\centering
\setlength{\tabcolsep}{2.5mm}
\renewcommand{\arraystretch}{1.2}

\caption{The experimental results between \tool and CodeT5 on false negativese.}
\resizebox{0.47\textwidth}{!}{
\begin{tabular}{c|c|ccc}

\toprule
\rowcolor[HTML]{DEDEDE}
\multicolumn{1}{c}{\textbf{Dataset}} & \multicolumn{1}{c}{\textbf{Baseline}} & \multicolumn{1}{c}{\textbf{Precision}} & \multicolumn{1}{c}{\textbf{Recall}} & \multicolumn{1}{c}{\textbf{F1 Score}} \\
\midrule
\multirow{2}{*}{PatchDB*}            & CodeT5                                & 73.62                                  & 58.52                               & 65.21                                 \\
                                     & \textbf{\tool}                      & \textbf{80.18}                         & \textbf{60.76}                      & \textbf{69.13}                        \\
\multirow{2}{*}{SPI-DB*}             & CodeT5                                & 61.17                                  & \textbf{62.59}                               & 61.87                                 \\
                                     & \textbf{\tool}                      & \textbf{78.07}                         & 61.79                      & \textbf{68.98}                               
\\
\bottomrule
\end{tabular}
}
\label{falsenegatives}
\end{table}

\subsection{Impact of Repository-level Context on False Negatives}

We also perform experiments on the impact of repository-level context on false negatives and the results are listed in Table~\ref{falsenegatives}. The experimental results demonstrate that \tool outperforms the current state-of-the-art baseline, CodeT5, in 5 out of 6 cases. Specifically, \tool achieves improvements of 6.56\% in precision and 2.24\% in recall on the PatchDB* dataset.
The sole exception is in the recall metric in SPI-DB*, where CodeT5 surpasses \tool by 0.8\%. It may be attributed to the additional context, which in some cases, has a minor impact on the rate of false negatives.

\subsection{Training and Inference Time}

\begin{table}[t]
\centering
\setlength{\tabcolsep}{2.5mm}
\renewcommand{\arraystretch}{1.2}

\caption{Time cost between \tool and CodeT5 per epoch training and inference time.}
\resizebox{0.47\textwidth}{!}{
\begin{tabular}{c|c|cc}

\toprule
\rowcolor[HTML]{DEDEDE}
\multicolumn{1}{c|}{\textbf{Baseline}} & \multicolumn{1}{c|}{\textbf{Time\textbackslash  Dataset}} & \multicolumn{1}{c}{\textbf{PatchDB*}} & \multicolumn{1}{c}{\textbf{SPI-DB*}}   \\
\midrule
\multirow{2}{*}{CodeT5}     & Train time (s)                            & 1573.56                  & 1171.95 \\
                                      & Inference time (s)                                       & 99.02                                 & 90.27   \\
\multirow{2}{*}{\tool}              & Train time (s)                                           & 420.45                                & 436.70  \\
                                      & Inference time (s)                                       & 37.20                                 & 27.38                                 
\\
\bottomrule
\end{tabular}
}
\label{timecost}
\end{table}

We have conducted a detailed analysis of the training and inference time costs per epoch for both \tool and CodeT5,  as presented in Table~\ref{timecost}.
The results indicate that \tool requires, only 857.15 seconds for training per epoch and 64.58 seconds for inference in the test set. In comparison, the best-performing baseline, CodeT5, requires 2745.51 seconds for training and 189.29 seconds for inference. The reason is that the graph branch is more efficient than the sequence branch (pre-trained model).

\subsection{Threats and Limitations}
We have identified the following major threats and limitations:

\textit{Constraints of Data Collection.}
To facilitate the identification of security patches at the repository level, we meticulously extracted 20,238 and
28,781 versions of repositories from platforms such as GitHub to reconstruct SPI-DB* and PatchDB*. Despite our extensive efforts to crawl the largest known repositories, some repositories listed in the National Vulnerability Database (NVD)~\cite{nvd} remained inaccessible. In our future work, we aim to expand our collection of security patches.

\textit{Generalizability on Other Programming Languages.} 
In this paper, we construct the \graph using Joern, Tree-sitter, and Cflow, specifically tailored for C/C++. While our experimental evaluations are focused on C/C++, the \tool can be extended to support other programming languages by integrating code analyzers that accommodate the respective syntax and structural paradigms. In future work, we plan to explore the applicability and effectiveness of \tool across wider programming languages, such as Java and Python, thereby broadening its applicability for security patch detection.

\textit{Restrictions of Dependency Extraction in Repository.}
We do not use the entire dependency graph in the repository due to the size of the \graph and the limitations inherent in static analysis techniques. While this approach is generally sufficient, it fails to encompass all scenarios, such as multi-level function calls within the repository. A potential solution to enhance the coverage could involve the development of additional rules and the application of more sophisticated slicing methods to construct the \graph.

\section{Related Work}\label{sec:related}
Security patch detection is crucial for enabling users to identify and apply updates addressing vulnerabilities on time~\cite{DBLP:conf/chi/VanieaR16}. It is important in OSS, where it is recommended to address vulnerabilities  silently until they are publicly disclosed~\cite{DBLP:conf/dsn/Wang0BJ19}. 
Initially, security patch detection primarily employed rule-based~\cite{DBLP:conf/sp/0002LTJ19, DBLP:conf/ndss/WuHML20} and traditional machine learning techniques~\cite{DBLP:conf/icse/TianLL12, DBLP:conf/uss/XuZZXB0L20, DBLP:conf/cns/WangW0BJ20, DBLP:conf/msr/SotoTWGL16, DBLP:conf/icse/CorleyKEL11}. For example, Li et al.~\cite{DBLP:conf/ccs/LiP17} conducted an empirical study on security patches. Wu et al.~\cite{DBLP:conf/ndss/WuHML20} developed symbolic rules to characterize security patches. VCCFinder~\cite{DBLP:conf/ccs/PerlD0AYRFA15/VCCFinder} utilizes SVM to identify potential vulnerabilities in open-source projects.
Subsequent research has incorporated DL-based methods for security patch detection~\cite{DBLP:conf/sigsoft/NguyenLKL022, DBLP:journals/corr/abs-2207-09022, DBLP:journals/tosem/ZhouSWLL22/SPIDB, DBLP:conf/icse/ZhouPCHXLH23}. Zuo et al.~\cite{DBLP:conf/sera/ZuoZSRF23} highlighted the role of commit messages in detecting security patches and introduced a Transformer-based approach. PatchRNN~\cite{DBLP:conf/milcom/WangWFSJBG21/patchrnn} integrates both source code and commit messages to improve performance for identifying security patches. GraphSPD~\cite{DBLP:conf/sp/WangWSJWL23/GraphSPD} proposes the PatchCPG and employs a graph-based approach for security detection. VulFixMiner~\cite{DBLP:conf/kbse/ZhouPW00WH21} extracts added and removed code from commits, utilizing CodeBERT to identify security patches in Java and Python.

Despite these advancements, existing methods do not consider the repository-level dependency and struggle to comprehend the relationship among multiple code changes within a patch. To address these challenges, 
we propose a repository-level security patch detection framework \tool.
We also extend both SPI-DB* and PatchDB* to the repository level.
\section{Conclusion}\label{sec:conclusion}
In this paper, we propose a repository-level security patch detection framework
named \tool, which comprises the \graph construction to incorporate repository-level dependency, a structure-aware patch representation for comprehending the relationship among multiple code changes, and progressive learning to facilitate the model in balancing semantic and structural information.
We further extend two widely-used datasets SPI-DB and PatchDB to the repository level, incorporating a total of 20,238 and 28,781 versions of repository in C/C++ programming languages, respectively (denoted as SPI-DB* and PatchDB*). Compared with the state-of-the-art approaches, the experimental results underscore the effectiveness of \tool for security patch detection.

Our source code and detailed experimental results are available at: \textit{{\http}}.

\bibliographystyle{IEEEtran}
\bibliography{sample-base.bib}

\end{document}